\documentclass[runningheads]{llncs}
\usepackage{multirow}
\usepackage{hyperref}

\usepackage{graphicx}
\usepackage{booktabs}
\usepackage{float}
\usepackage{amsmath,amsfonts,amssymb}
\usepackage{color}
\usepackage{wrapfig}
\usepackage{subcaption}
\usepackage[misc]{ifsym}
\makeatletter
\newcommand{\printfnsymbol}[1]{%
  \textsuperscript{\@fnsymbol{#1}}%
}

\begin{document}
\title{SIMBA: Specific Identity Markers for Bone Age Assessment}
%
%
\author{Cristina Gonz\'alez\inst{1}\thanks{Both authors contributed equally to this work.}(\Letter) \and
Mar\'ia Escobar\inst{1}\printfnsymbol{1}\and
Laura Daza\inst{1} \and
Felipe Torres\inst{1} \and
Gustavo Triana\inst{2} \and
Pablo Arbel\'aez\inst{1}}

%
%
\authorrunning{C. Gonz\'alez et al.} 

\institute{Center for Research and Formation in Artificial Intelligence,\\
Universidad de los Andes, Bogot\'a, Colombia \\
\email{\{ci.gonzalez10, mc.escobar11, la.daza10, f.torres11, pa.arbelaez\}@uniandes.edu.co} \and
Fundaci\'on Santa Fe de Bogot\'a, Bogot\'a, Colombia}
\maketitle              
%
\begin{abstract}
 Bone Age Assessment (BAA) is a task performed by radiologists to diagnose abnormal growth in a child. In manual approaches, radiologists take into account different \textit{identity markers} when calculating bone age, i.e., chronological age and gender. However, the current automated Bone Age Assessment methods do not completely exploit the information present in the patient's metadata. With this lack of available methods as motivation, we present SIMBA: Specific Identity Markers for Bone Age Assessment. SIMBA is a novel approach for the task of BAA based on the use of identity markers.
 For this purpose, we build upon the state-of-the-art model, fusing the information present in the identity markers with the visual features created from the original hand radiograph. We then use this robust representation to estimate the patient's relative bone age: the difference between chronological age and bone age. We validate SIMBA on the Radiological Hand Pose Estimation dataset and find that it outperforms previous state-of-the-art methods. SIMBA sets a trend of a new wave of Computer-aided Diagnosis methods that incorporate all of the data that is available regarding a patient. To promote further research in this area and ensure reproducibility  we will provide the source code as well as the pre-trained models of SIMBA. 

\end{abstract}

 \section{Introduction}
 The height of a child is one of the best indicators for general health and overall well-being. Early diagnosis of abnormal growth in children is relevant not only for predicting the final adult height but also for detecting potential endocrine disorders. Prior studies have found that early recognition of abnormal growth in children is necessary for timely treatment of pathological conditions such as precocious puberty \cite{haymond2013early,oerter1999precocious}. Physicians evaluate the growth rate of a child through Bone Age Assessment (BAA), a measurement of a child's skeletal development in months that varies according to \textit{identity markers}, such as the child's chronological age, gender, and ethnicity.
 
Currently, the way in which radiologists establish a child's bone age is by comparing the child’s hand radiograph against atlases for bone age measurement. These atlases contain standard reference images portraying male and female bone development from birth to an estimate of the last years of bone development for each gender. In the Greulich and Pyle (G \& P)\cite{gp} atlas, the physician finds the reference image that presents the most similarities with the hand radiograph of the patient and uses it as a guideline to determine bone age. Tanner and Whitehouse's (TW2) \cite{tanner} method is based on identifying anatomical Regions of Interest (RoIs) on the epiphysis of the hand and wrist, assigning a score to each RoI and combining them to calculate an overall bone age. In both approaches, the radiologist considers the patient's specific identity markers, particularly the gender and chronological age. 

\begin{figure}[t]
   \centering
   \medskip
   \begin{subfigure}[t]{.3\linewidth}
     \centering\includegraphics[width=\linewidth]{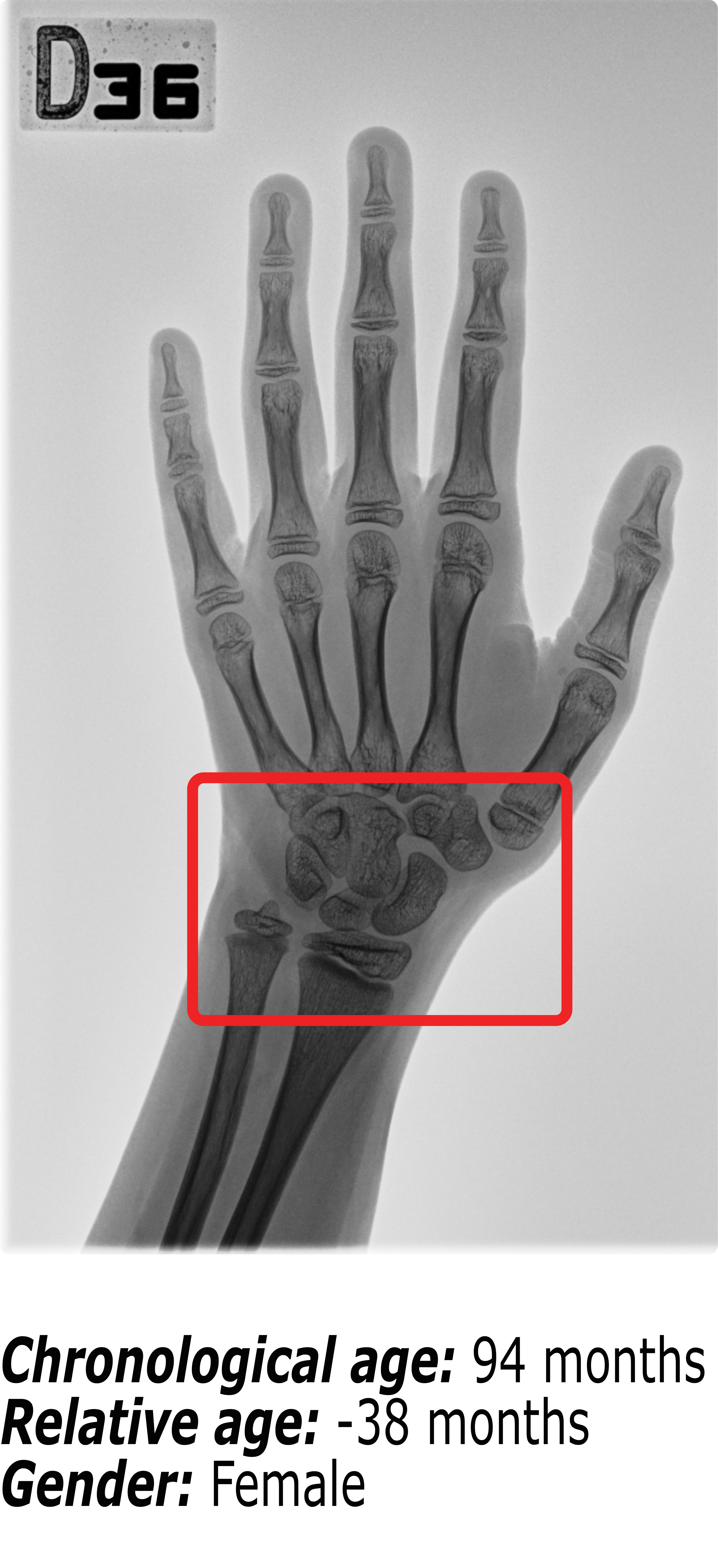}
     \subcaption{}
     \label{fig:age_1}
   \end{subfigure}
   \begin{subfigure}[t]{.3\linewidth}
     \centering\includegraphics[width=\linewidth]{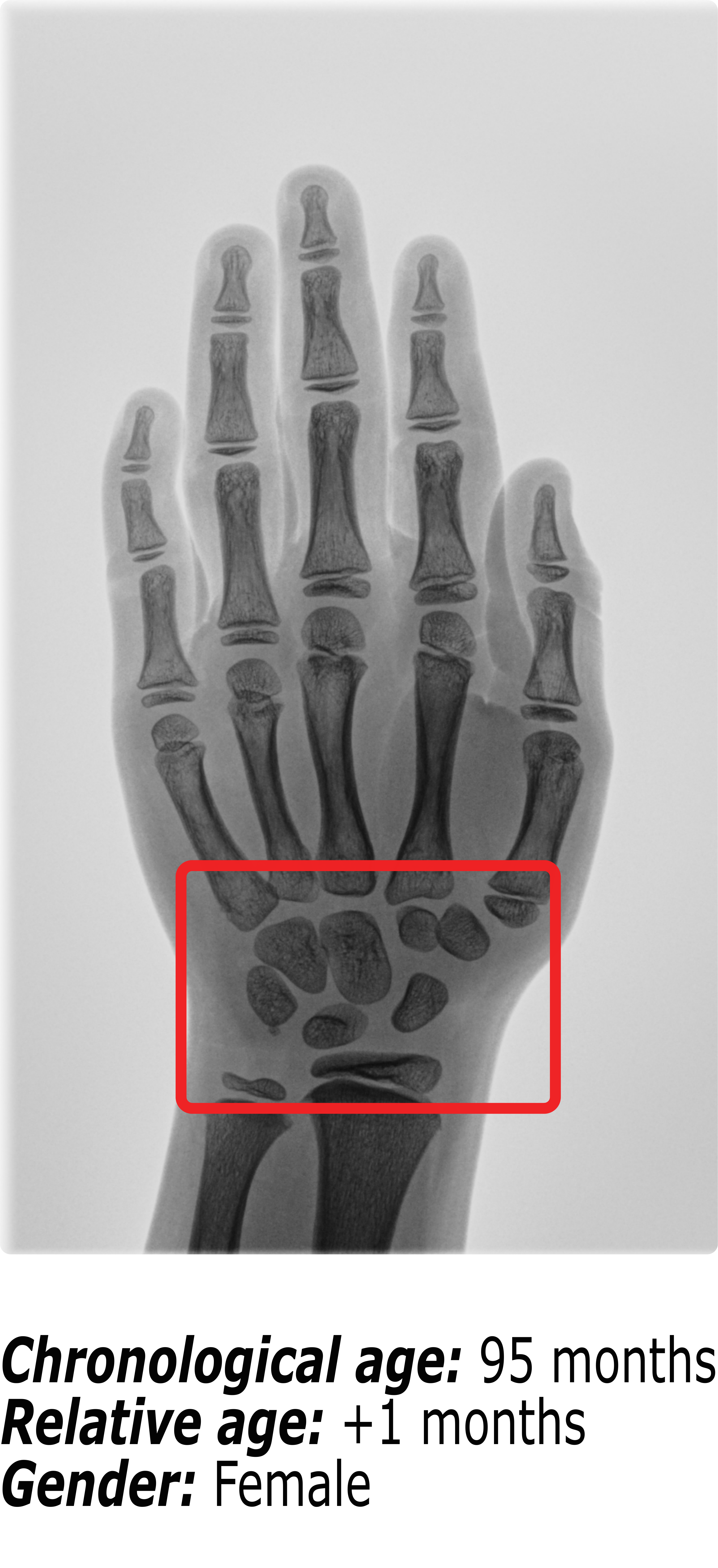}
     \subcaption{}
     \label{fig:age_2}
   \end{subfigure}
   \begin{subfigure}[t]{.3\linewidth}
     \centering\includegraphics[width=\linewidth]{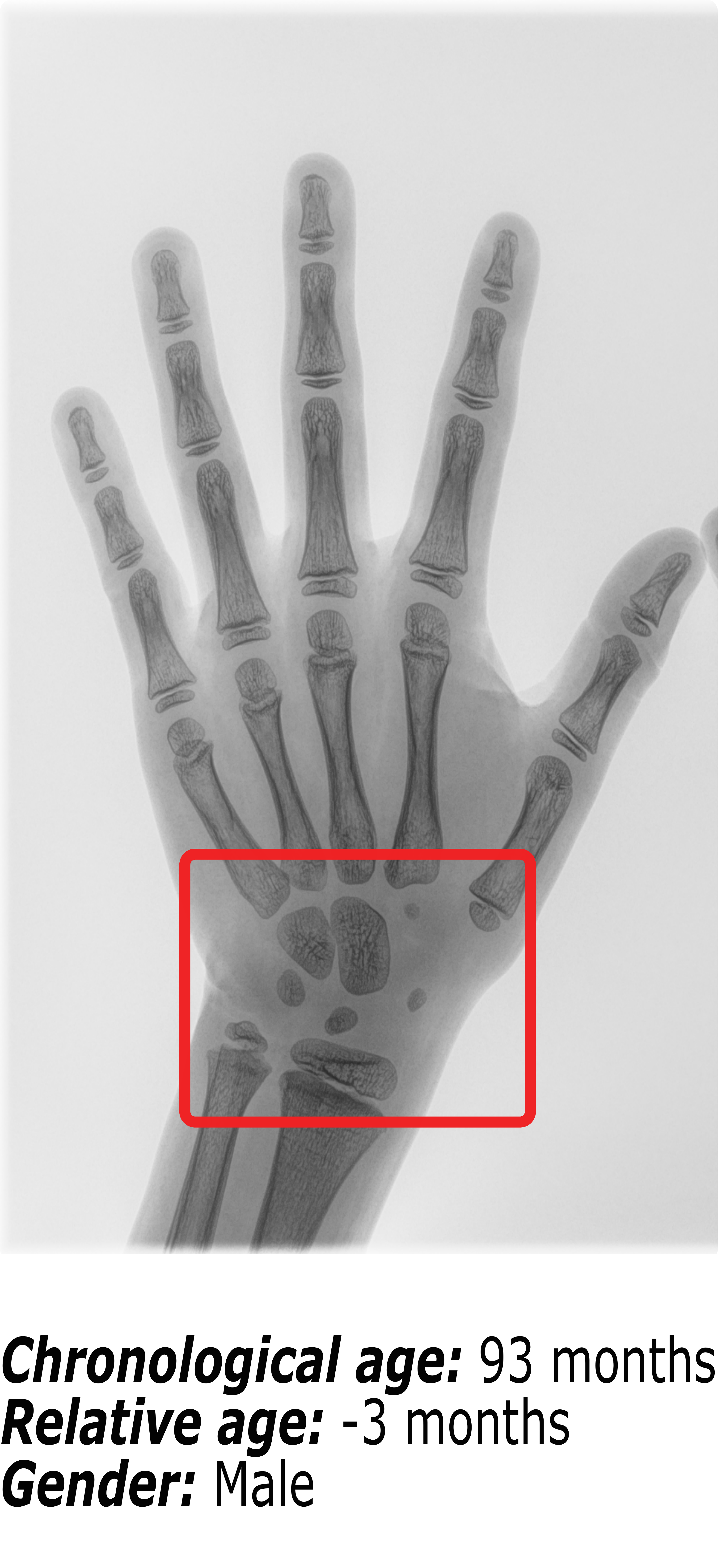}
     \subcaption{}
     \label{fig:age_3}
   \end{subfigure}
   \caption{Hand radiograph of three patients with virtually the same chronological age. Despite the small variations in chronological age, the anatomical structures, surrounded by the red box, have a very diverse appearance due to differences in the ossification patterns. These difference are related to the relative bone age of the patient with respect to the chronological age and their gender.}
  \label{fig:dataset_ages}
 \end{figure}

Figure \ref{fig:dataset_ages} shows an example of the influence that \textit{identity markers} have on bone age. The three hand radiographs present in Figure \ref{fig:dataset_ages} belong to children with virtually the same chronological age. However, it is visible that the ossification patterns present in each of the hand radiographs vary significantly. First, gender is an important identity marker to take into account. Comparing Figure \ref{fig:age_2} (a female patient) and Figure \ref{fig:age_3} (a male patient), it is possible to observe that the bone structures in the region surrounded by the red box are more developed for the female than for the male. This finding is supported by the fact that skeletal development is faster in females than in males. 

Nonetheless, when comparing two hand radiographs of patients of the same gender  with the same chronological age, as it happens between Figure \ref{fig:age_1} and Figure \ref{fig:age_2}, the expected result would be that the bone patterns did not vary much. Because most patients have a regular growth pattern, physicians use the chronological age as a starting point and compute the difference in skeletal development. This \textit{relative bone age} between the patient's chronological age and the patient's bone age is the information that the radiologists use to diagnose growth disorders. However, Figure \ref{fig:age_2} belongs to a patient with regular growth, having a relative bone age of +1 month, while Figure \ref{fig:age_1} belongs to a patient with accelerated growth, hence the relative bone age of -38 months. 

Since physicians have to take into account different factors before determining bone age, BAA is highly dependent on the radiologist's expertise level. Automated methods for BAA have been proposed as an alternative for manual approaches in order to reduce the variability among radiologists. The only commercial automated method is BoneXpert \cite{bonexpert}, a private software based on edge detection and active appearance models \cite{Cootes98activeappearance} currently used in clinical settings. However, the algorithm was developed using patients from a single cohort. Therefore there is no guarantee that it can generalize the BAA for children with different \textit{identity markers} from those of the patients with whom the model was trained. Like BoneXpert, the first wave of automated BAA methods \cite{Tsao2008,Liu2008} and digital atlases \cite{gaskin2011skeletal,gertych2007bone,gilsanz2005hand} were private. Limiting the comparison that could be done among methods and the disposition that researchers had in choosing BAA as a relevant problem to tackle. 

To motivate the development of more general automatic BAA methods, the Radiological Society of North America (RSNA) organized a challenge in 2017 with a dataset containing patients from different hospitals \cite{rsna_pediatric_boneage_challenge_2017}. The winners of this challenge, 16 bit \cite{met2}, developed a method that uses global information of the hand radiograph image and handcrafted embedding for gender. In our previous work \cite{bonet}, we created the Radiological Hand Pose Estimation (RHPE) dataset, which includes information of bone age, gender, anatomical RoIs, and chronological age for a cohort with a different ethnicity than the previously available dataset. There have been several automated methods for BAA that focus on a global approach \cite{rsna_pediatric_boneage_challenge_2017,larson,pan2020fully,torres2020empirical}, like what physicians do in G \& P, and other approaches that exploit the local information \cite{met2,bonet,met1,prsnet,liu}, in a way inspired by TW2, to predict the bone age of the child.

\begin{figure*}[t]
\centering
\includegraphics[width=1.0\linewidth]{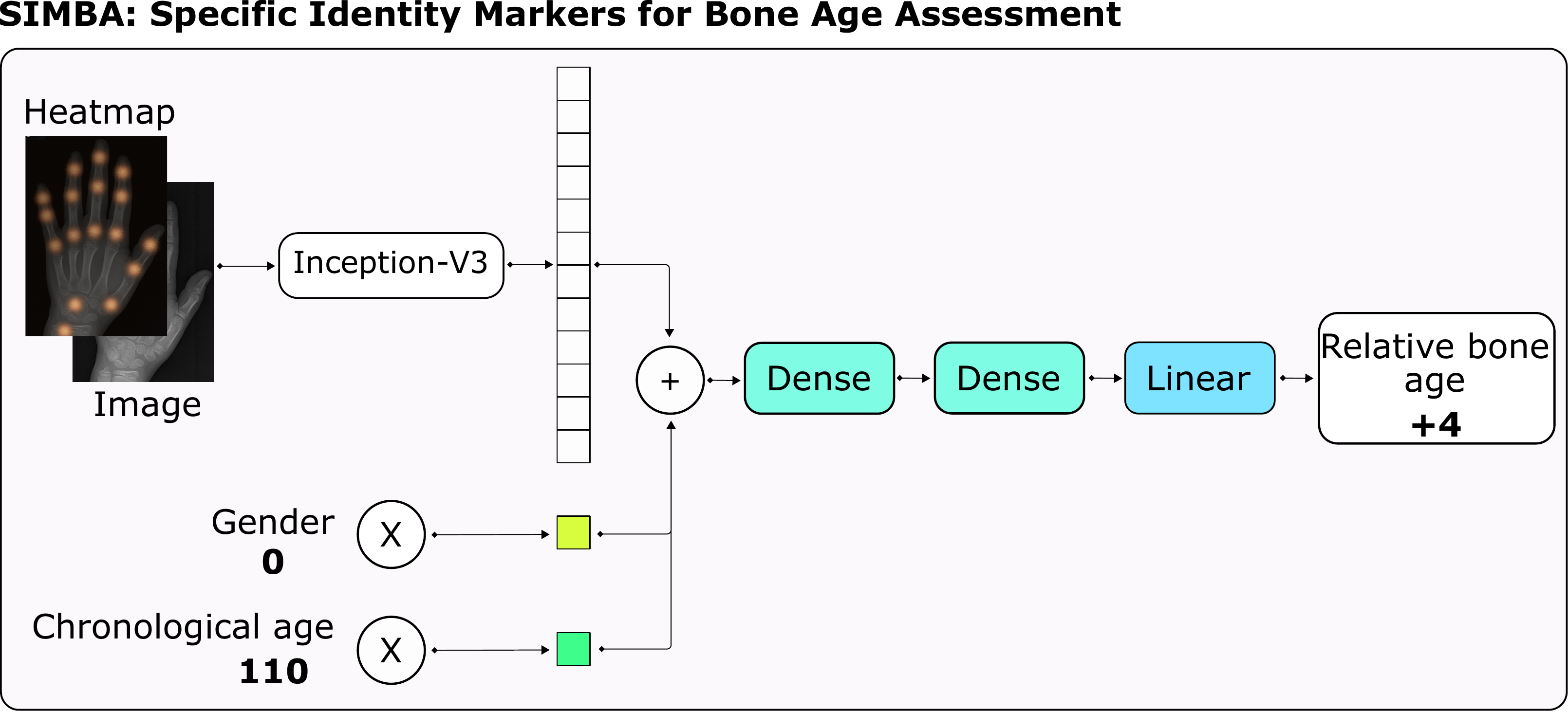}
\caption{Overview of our method. Our model takes as input the hand radiograph and the attention heatmap generated from the anatomical RoIs. We calculate visual features from these two inputs using an Inception-V3 architecture. We multiply the identity markers with two learnable and independent multipliers and concatenate this information with the visual features extracted. Finally, we process all this information, the visual features and the identity markers, jointly to predict the bone age deviation from the chronological age. Best viewed in color}
\label{fig:overview}
\end{figure*}

In this paper, we present Specific Identity Markers for Bone age Assessment (SIMBA). Figure \ref{fig:overview} shows an overview of our method. Motivated by the way physicians estimate bone age in children, SIMBA builds upon the state-of-the-art method, BoNet \cite{bonet}, and incorporates patient-specific identity markers, \textit{i.e.} chronological age and gender, to perform BAA. State-of-the-art methods that use gender information for BAA introduce this input to the model both, directly as an additional input\cite{met2,prsnet}, and indirectly training an additional model \cite{rsna_pediatric_boneage_challenge_2017}. Experimentally, we demonstrate that our way of incorporating gender information is more effective for BAA, as our model significantly outperforms the state-of-the-art methods in the RHPE dataset. We extract high-level features from the hand radiograph implicitly guided by an attention heatmap over the anatomical RoIs, following the idea suggested by \cite{bonet}. Our model then uses learnable independent multipliers for each identity marker and combines them with the image features to generate the prediction. We also introduce relative bone age as a new way of approaching the problem of BAA in a fashion similar to physicians. We evaluate SIMBA on the open source RHPE dataset, outperforming the current state-of-the-art.

 Our main contributions can be summarized as follows:
\begin{enumerate}
    \item We propose a novel way to incorporate identity markers into BAA methods. We demonstrate that using a patient's gender and chronological age as prior for the model is relevant for better BAA.
    \item We demonstrate that addressing the problem of BAA by estimating the relative bone age with the prior of chronological age is relevant for better BAA.
 \end{enumerate}
 
In order to ensure the reproducibility of our results and promote further research on BAA, we provide the pre-trained models, the source code for SIMBA and the additional metadata corresponding to the chronological age for the RHPE dataset.\footnote{\url{https://github.com/BCV-Uniandes/SIMBA}}

\section{SIMBA}
Our method is inspired by how radiologists take advantage of all the available information for each patient when computing BAA. Thus, SIMBA not only considers visual information from the hand radiograph and local information from anatomical RoIs, but it also leverages identity markers, particularly, chronological age and gender. Additionally, we propose a novel paradigm of predicting the relative bone age of a patient as the deviation from the chronological age. Figure \ref{fig:overview} depicts our approach for incorporating a patient’s specific identity markers to predict relative bone age. In the following sections, we explain the different components of SIMBA in more detail.
 
 \subsection{Specific Identity Markers}
Similarly to the state-of-the-art method BoNet, given an image $I$ and a heatmap $H$ our architecture incorporates the Inception-V3\cite{inception} ($I_{V3}$) network to extract visual features, $\mathbb{I}$.

\begin{equation*}
    \mathbb{I} = I_{V3}(I, H)
\end{equation*}

Among a patient’s identity markers, the most relevant ones for BAA are gender and chronological age, due to skeletal development varying with gender and being correlated to chronological age. We incorporate these identity markers directly by processing them jointly with the previously extracted visual features. Instead of using handcrafted embeddings for incrementing dimensionality \cite{bonet}, we learn multipliers $m_{g}$ and $m_{c}$ to balance the importance of each of the inputs from which our model makes the final prediction, regardless of their size. Therefore, our model learns weighted representations for gender $\mathbb{G}$ and chronological age $\mathbb{C}$, according to their relevance to the final prediction. These representations given a gender $g$ and a chronological age $c$ are defined as follows:

\begin{equation*}   
\mathbb{G} = m_{g} \cdot g
\end{equation*}
\begin{equation*}
    \mathbb{C} = m_{c} \cdot c
\end{equation*}

Thus, the joint final representation $\mathbb{J}$ ,which our model uses to estimate the child's bone age, corresponds to the concatenation of the visual features extracted from the image and the heatmap, along with the weighted representations of the gender and chronological age.
\begin{equation*}
    \mathbb{J} = [\mathbb{I}; \mathbb{G}; \mathbb{C}]
\end{equation*}

 \subsection{Relative bone age}
 We propose a paradigm shift in terms of the formulation of the BAA task. We define a new task for BAA equivalent to the one previously formulated for this problem. Based on the priors it receives, our model is optimized to predict the difference between the chronological age $c$ and the bone age $b$ of the patient, defined as the relative bone age $r_{b}$.  In other words, our model learns to take as input a chronological age and outputs a residual bone age.
\begin{equation*}
    r_{b} = c - b
\end{equation*}

For this purpose, our model learns two intermediate representations from linear layers followed by $ReLU$ non-linear activation function ($Dense$). From these layers, the model learns to generate a joint representation $\widehat{\mathbb{J}}$ of the visual information and the specific identity markers of the patient. Finally, SIMBA predicts the relative bone age with a fully-connected layer.

\begin{equation*}
    Dense(x) = ReLU(W(x) + b)
\end{equation*}
\begin{equation*}
    \widehat{ \mathbb{J}} = Dense(Dense(x))
\end{equation*}
\begin{equation*}
    r_{b} = W\widehat{\mathbb{J}} + b
\end{equation*}

 \textit{Implementation details}: we train our method on an NVIDIA TITAN-X Pascal GPU for 150 epochs with an initial learning rate of $0.001$, 17 images per batch and use an Adam \cite{adam} optimizer with the standard parameters. Additionally, we use a dynamic learning rate scheduler to reduce the learning rate when reaching a plateau with a patience of 2 epochs, a reducing factor of 0.8, and a cooldown of 5 epochs.
 
\section{Experiments}
\subsection{Experimental setup}
For our experimental validation, we use the RHPE dataset with the original data splits. We perform an ablation study to determine the individual contribution of each module. For the ablation experiments, we train our method on the training set and select the model that performs better on the validation set. Additionally, for the comparison of our method with respect to the state-of-the-art, we train our best model using the data from both the training and validation set and evaluate on the official RHPE test server.

In accordance with our new formulation of the task, we aim at estimating the deviation of the bone age from the chronological age of the patient in months for each given image in the dataset. To evaluate our experimental results, we rely on the Mean Absolute Distance (MAD) previously used in the RSNA 2017 Pediatric Bone Age Challenge \cite{rsna_pediatric_boneage_challenge_2017}.

\subsection{Experimental validation}
\subsubsection{Comparison with the state-of-the-art}
We compare the results of our method, SIMBA, with respect to the state-of-the-art methods in this dataset. Table \ref{tab:soa} shows results in the test set of the RHPE dataset for the methods as reported by \cite{bonet}, 16 bit and BoNet. 

\begin{table}[t]
\centering
\caption{Comparison of our method against the state-of-the-art methods, as reported on \cite{bonet}, on the RHPE test set. Our method SIMBA significantly outperforms the state-of-the-art-methods.}
\label{tab:soa}
\begin{tabular}{cc}
\toprule
Method & MAD \\
\midrule
16 bit\cite{met2} & 8.57 \\
BoNet\cite{bonet} & 7.60 \\
SIMBA (Ours) & \textbf{5.47} \\
\bottomrule
\end{tabular}%
\end{table}

The results reported in Table \ref{tab:soa} demonstrate that SIMBA significantly outperforms the state-of-the-art method in the RHPE dataset test set. Since the other state-of-the-art methods do not include the chronological age information as prior for the model, our results empirically demonstrate that including identity markers, specifically the gender and chronological age of the patient, is important to improve performance in BAA. Additionally, we demonstrate that when we replace the task of estimating bone age with estimating its deviation with respect to chronological age, it is evident in performance improvements for BAA.
\subsubsection{Relative age bias analysis}
\begin{figure*}[!t]
\centering
f\includegraphics[width=0.75\linewidth]{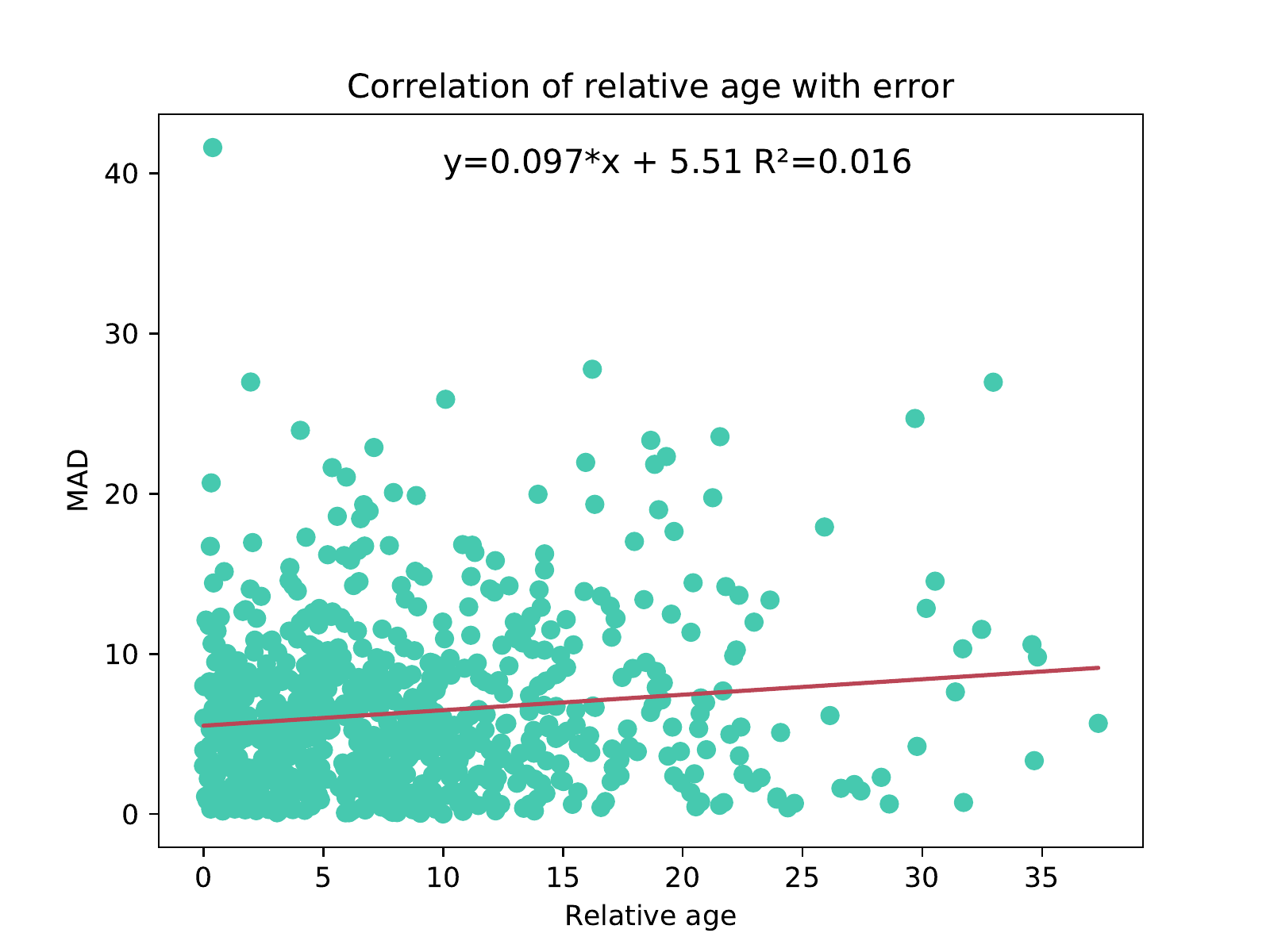}
\caption{Correlation between relative age and MAD for the validation set of RHPE. There is no strong linear dependence (the pink line has a tendency to be horizontal), therefore SIMBA is not biased towards relative age. Best viewed in color}
\label{fig:correlation}
\end{figure*}

To gain further insight, we calculated the correlation between the relative age and the MAD metric for all the patients in the validation split. As shown in Figure \ref{fig:correlation}, we estimated the correlation coefficient to measure the linear relationship between these variables. The correlation coefficient is 0.016, and thus, we can state that there is no strong linear dependence. Furthermore, we performed a linear regression on the data and found that the slope of the line is 0.097, which is consistent with an approximately uniform distribution of MAD with respect to relative age. Based on this analysis, we can conclude that SIMBA is not biased towards relative age and learns to predict a residual bone age based entirely on the visual input and the guidance of the identity markers.

\subsubsection{Ablation study}
We designed an ablation study of our method for its different components. For all our experiments we train our model incrementally starting from the baseline. We use the source code publicly available for BoNet. In this way, to build the final model, we add our modifications in the following order: gender multiplier, chronological age, and relative bone age, according to the components included in each ablation experiment. Table \ref{tab:ablations} shows the results of our final model, the ablation experiments, and the baseline in the validation split of the RHPE dataset.

\begin{table}[!b]
\centering
\caption{Ablation experiments for our method on the RHPE validation set. We report the results of our final method, ablation experiments for: relative bone age, chronological age and gender multiplier, and the baseline of our experimental setup.}
\label{tab:ablations}
\begin{tabular}{ccccc}
\toprule
\multirow{2}{*}{Method} & \multicolumn{2}{c}{Identity markers} & \multirow{2}{*}{Relative bone age} & \multirow{2}{*}{MAD} \\
 \cmidrule{2-3}
 & Gender & Chronological age &  &  \\
  \midrule
BoNet\cite{bonet} & & & & 7.48 \\
 \midrule
 \multirow{4}{*}{SIMBA} & \checkmark & \checkmark &  & 6.50 \\
 & \checkmark &  & \checkmark & 8.72 \\
 &  & \checkmark & \checkmark & 7.33 \\
 & \checkmark & \checkmark & \checkmark & \textbf{6.34}\\
 \bottomrule
\end{tabular}%

\end{table}
\textit{Relative bone age ablation} If we train our model without establishing the final task as the estimation of the deviation of the patient's bone age with respect to their chronological age, the error of our model increases by 0.16 months in the validation set of the RHPE dataset. These results demonstrate that the task that we propose allows our model to be able to exploit the input information, that is, the image and the identity markers more adequately.\\

\textit{Chronological age ablation}. By eliminating the chronological age as a prior for our model, the MAD increases by 2.38 months. This result empirically supports that the decrease in error associated with the introduction of the task of the estimation of relative bone age is determined by having the prior of chronological age. We consider the above to be intuitive if we understand that, by not including the chronological age, the model must learn to estimate not only the bone age but also the chronological age of the patient to finally estimate the deviation between them. However, this does not represent the real case since the radiologist usually knows the patient's chronological age.\\

\textit{Gender multiplier ablation}. If we change our gender multiplier for the handcrafted embedding used in the state-of-the-art methods for BAA, the MAD increases by 0.99 months. This result shows that our multiplier exploits gender information more efficiently. Additionally, the effectiveness of the other contributions of our model are highly related to the joint processing of the identity markers of the patient.\\

\section{Conclusions}
In this work, we present a new paradigm for the task of BAA by estimating the deviation between the bone age and the chronological age of a patient. To the best of our knowledge, SIMBA is the first method for this task that leverages information from specific identity markers, particularly the gender and the chronological age of a patient. Our model outperforms the state-of-the-art method in the test set of the benchmark RHPE dataset. The previous state-of-the-art methods do not consider chronological age as an identity marker and directly estimate the child's bone age. We demonstrate experimentally that including prior information related to specific identity markers of the patient into the network, inspired by the way radiologists do it in their medical practice, results in a more accurate Bone Age Assessment.

\paragraph{Acknowledgments}
This project was partially funded by  the Colombian Ministry of Science, Technology and Innovation under the Colciencias grant: 841-2017 code 120477758362.

%

\end{document}